\documentclass[prl,twocolumn,superscriptaddress,showpacs,floatfix]{revtex4}

\usepackage{times}
\usepackage{amsmath}
\usepackage{amssymb}
\usepackage{graphicx}
\usepackage[noxcolor]{pstricks}

\newcommand{\ZZ}{\mathbb{Z}}

\begin{document}



\title{A theory of topological edges and domain walls}

\author{F.A.~Bais}
\affiliation{Institute for Theoretical Physics, University of Amsterdam, Valckenierstraat 65, 1018 XE Amsterdam, The Netherlands}
\affiliation{Santa Fe Institute, Santa Fe, NM 87501, USA}

\author{J.K.~Slingerland}
\affiliation{Dublin Institute for Advanced Studies, School of Theoretical Physics, 10 Burlington Rd, Dublin, Ireland}
\affiliation{Department of Mathematical Physics, National University of Ireland, Maynooth, Ireland}

\author{S.M.~Haaker}
\affiliation{Institute for Theoretical Physics, University of Amsterdam, Valckenierstraat 65, 1018 XE Amsterdam, The Netherlands}


\begin{abstract}
We investigate 
domain walls between topologically ordered phases in two spatial dimensions and present a simple but general framework from which their degrees of freedom can be understood. The approach we present exploits the results on topological symmetry breaking that we have introduced and presented elsewhere. 
After summarizing the method, we work out predictions for the spectrum of edge excitations and for the transport through edges in some representative examples. These include domain walls between the Abelian and non-Abelian topological phases of Kitaev's honeycomb lattice model in a magnetic field, as well as recently proposed domain walls between spin polarized and unpolarized non-Abelian fractional quantum Hall states at different filling fractions.
\end{abstract}

\date{\today}


\maketitle

\section{Introduction}

Recently there has been considerable interest in planar systems which exhibit topological phases. These phases are characterized by topological field theories (TQFTs) or corresponding conformal field theories (CFTs). It is of great interest to have a clear understanding of the  edges of such systems, and of domain walls between regions in different phases. In fractional quantum Hall systems, where experimental support for the existence of a variety of topological phases is strongest, observations are almost entirely restricted to edge transport, and proposed devices for probing the topological order rely on interference of tunneling currents between edges \cite{DasSarma05,Nayak2007,Bonderson07b}. In such experiments the electron density is usually not constant throughout the sample and islands with different filling fractions form, separated by domain walls. In lattice models with several topological phases, one may similarly induce phase boundaries by varying the local couplings.



In this letter, we present a general method to determine the degrees of freedom of boundaries between topological phases and their relation to the bulk degrees of freedom, based on the condensation of bosonic quasiparticles in auxiliary layered systems.
Consider a medium in topological phase I supporting a single island in topological phase II. 
A setup like this has been considered in experiments by Camino et al.~\cite{Camino05a, Camino05b} for the case where phase I is the fractional quantum Hall state at filling fraction $\nu=1/3$ and phase II the FQH state at $\nu=2/5$. We plan to give a treatment of this situation and more generally, of the hierarchy states described in \cite{haldane,halperin} and \cite{Bonderson07c} in a separate publication \cite{BaisWIP}.  Here, we describe how the topological symmetry breaking procedure of \cite{BSS02,BSS03,baismathy06b,Bais2008-08} can be applied in such physical settings, and follow up with two different examples. These involve Kitaev's spin model on the honeycomb lattice~\cite{Kitaev06a} and a domain wall between spin polarized and unpolarized non-Abelian fractional quantum Hall liquids~\cite{Grosfeld2008}.

\section{Walls from topological symmetry breaking}

\begin{figure}[htbp] 
   \centering
   \includegraphics[width=3in]{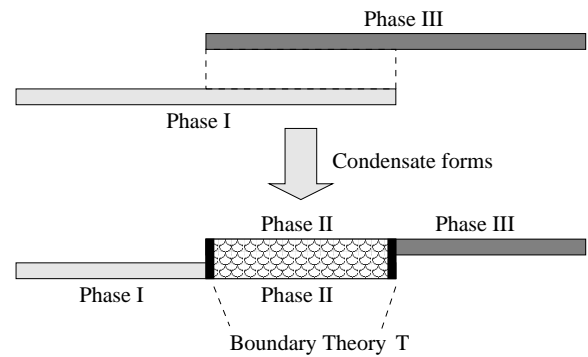} 
   \caption{Side view of two overlapping layers supporting topological phases I and III. If we bring the layers close together, a condensate may form in the overlap region leading to a phase II. The theory T on the left boundary describes excitations that can be divided into bulk excitations of phase I and of phase II, and excitations that only can propagate along the boundary. On the right boundary a similar situation occurs for the same theory T, now with III replacing I. The subset of T excitations that are strictly confined to the left and right boundaries are therefore different in general.}
   \label{fig:hallbars}
\end{figure}

One way to match two different phases on a boundary is by taking the tensor product of the two boundary CFTs. This would correspond to the situation where the two phases are separated by a strip of vacuum. 
To describe more general interfaces, we consider the geometry sketched in figure \ref{fig:hallbars}, where we started with two layers in phases I and III, which we let partially overlap as indicated. In the middle section we have a two layer system. 
If we bring the layers close we may have some binding between degrees of freedom in I and III. In particular, a bosonic composite of excitations from the two layers could occur, and consequently, a condensate of such bosons may form. This condensation will lead to a different phase for the middle region, which we denote by II. 


This situation can be analysed with the tools we have developed in \cite{BSS02,BSS03,baismathy06b,Bais2008-08}. Let us assume we are given CFTs or TQFTs $C_1$ and $C_3$ describing phases I and III, or even just the spectrum of topological sectors $a,b,\ldots$, the fusion rules and the topological spins $\theta_{a}=e^{2i\pi h_{a}}$ for these phases (the topological spins $\theta_{a}$ correspond to the fractional parts of the conformal weights $h_{a}$). The topological sectors of the $C_1\otimes C_3$ theory will be labeled by pairs $(a^{I},b^{III})$ of labels from the $C_1$ and $C_3$ theories. We can now proceed as follows. 

\noindent (i) We can find out which bosonic sectors are possible in the $C_1\otimes C_3$ theory. Bosons always have $\theta_{(a,b)}\equiv\theta_{a}\theta_{b}=1$ and in the simple case of sectors with quantum dimension $d_{(a,b)}\equiv d_a d_b=1$, this requirement is all that is needed. 

\noindent (ii) We assume that a condensate forms in one or more of the bosonic channels and use our method \cite{Bais2008-08} to determine the topological spectrum and fusion rules of the condensed phase. We denote the residual theory by $T$. Sectors of the $C_1\otimes C_3$ theory branch into \mbox{$T$-sectors} according to branching rules of the form $(a^{I},b^{III})\rightarrow \sum_{c} N_{(a,b)}^{c} c^{T}$, where the $N_{(a,b)}^{c}$ are integer branching multiplicities. Thus in the new phase, some sectors, or primary fields of the $C_1 \otimes C_3$ theory that we started off with may branch to the same unique $T$-sector and hence will be identified, while others may split into independent $T$-sectors. Typically, sectors which are related by fusion with the condensed boson are identified, while sectors which are invariant under fusion with the condensed boson may split. The condensed sector itself always branches to the vacuum of the $T$-theory. We require that branching and fusion are compatible, i.e. they commute. As a result, branching conserves quantum dimensions.  

\noindent (iii) While the $T$-theory is required to have good fusion rules, some $T$-sectors will not inherit well defined spin factors or conformal weights from the uncondensed theory, basically because they have nontrivial braiding interaction with the condensed excitation. The corresponding excitations will pull strings in the condensed medium and will be confined. In effect, this means that they are expelled from the bulk and can propagate only on the boundary of the condensed medium. 

\noindent (iv) The $T$-sectors which do inherit well defined topological spins from the uncondensed theory survive in the bulk and define the theory $C_2$, which is a TQFT which describes the fusion and braiding of excitations of phase II.

We now have a clear picture of the situation before and after condensation. Before condensation, arbitrary excitations of the system could be labeled $(a^{I},b^{III})$, with excitations in phases I resp.~III labeled $(a^{I},1)$ and $(1,b^{III})$ resp.,~where $1$ denotes the vacuum, or topologically trivial sector. (We may use this notation even where the second layer is not present.) After condensation, excitations in phases I and III are labeled as before, but in the overlap region, we now have phase II, with bulk excitations described by the unconfined $T$-sectors and boundary excitations described by the confined $T$-sectors. 

Using the fusion rules of the $T$-theory, we can understand all the kinematics of processes that may occur when excitations are moved toward or through the edges. The crucial insight here is that excitations in all parts of the system can be labeled by sectors of the $T$-theory. For the bulk and boundary of phase II this is true by definition and for the bulk of phase I we can use the branching rules to obtain the possible $T$-sectors from the actual sectors $(a^{I},1)$. The situation is particularly simple if every sector of phase I branches to a unique $T$-sector. This happens under quite general circumstances, notably whenever the theory $C_1$ does not by itself have bosonic sectors. We can now classify all boundary processes that may occur. For example any $C_1$-particle that is identified with a non confined $T$-particle can pass through the phase boundary unnoticed and vice versa, while  a $C_1$-particle that corresponds to a confined $T$-particle cannot enter the region in phase II. On the other hand, $T$-particles which are confined in phase II but which can be obtained from a $C_1$ sector by branching can pass into the area in phase I after being driven out of phase II, without leaving a trace on the boundary. Hence the true boundary excitations are labeled by the confined $T$-sectors which do not correspond to $C_1$-sectors. For processes involving three or more excitations, we need to use the fusion rules of $T$. Any process allowed by these rules could in principle occur. For example, a $C_1$ particle corresponding to a confined $T$-sector $c$ could hit the phase boundary and split into a boundary excitation $a$ and a bulk excitation $b$ of phase II, provided that $c\in a\times b$ according to the fusion rules of $T$. 

The fusion rules of $T$ are valid throughout. For instance, if two particles in phase I have a definite joint fusion channel, then this should be preserved even if one of the particles is moved into the region in phase II. In fact, the full topological state of this multi-phase system should be characterized by specifying the amplitudes for the $T$-fusion channels obtained on successive fusions of all the quasiparticles that are present.  This yields fusion bases analogous to the standard fusion bases of the the topological Hilbert spaces of systems described by a single TQFT. Clearly, to actually perform the fusions involved, it will usually be necessary to bring the quasiparticles from the bulk of phases I and II to the boundary.

There are many situations where these general ideas apply. In particular, coset models form a large class that can be analyzed in terms of bose condensates \cite{Bais2008-08}. The construction of these models closely parallels the construction of figure \ref{fig:hallbars}, where  phase I is a $G_{k}$ phase, phase III is a $H_{k'}$ phase with the opposite chirality and in the overlap region we obtain a phase with the topological order of the $G_{k}/H_{k'}$ coset, after condensation of all available bosons. We continue with two concrete applications of a slightly different, but related type.

\section{Kitaev's honeycomb and the toric code}

Kitaev's honeycomb model~\cite{Kitaev06a} is a model of spins living on the sites of a honeycomb lattice and interacting through nearest neighbor Ising-like interactions. The model is exactly solvable and displays two types of phases: 
three equivalent gapped Abelian topological phases with the same topological order as the $\ZZ_2$ toric code model and central charge $c=0$, and a gapless phase, which becomes gapped when a Zeeman term is added to the Hamiltonian and then displays non-Abelian topological order described by the Ising TQFT at $c=1/2$.  
The Abelian phase has four sectors with $\ZZ_2\times \ZZ_2$ fusion rules and the Ising model has the well known three sectors labeled $1$, $\sigma$, $\psi$, with $1$ denoting the vacuum and with nontrivial fusion rules given by $\sigma\times\sigma = 1+\psi$, $\sigma\times\psi=\sigma$ and $\psi\times\psi=1$. 
\begin{table}[htb]
\begin{center}
$\begin{array}{l|l|l|l}
\multicolumn{4}{l}{Ising \vphantom{\ZZ_2}}\\
\hline
 c= 1/2 & 1 & \sigma& \psi \\
 \hline
 h_i & 0 & \frac{1}{16}& \frac{1}{2} \\
 d_i & 1 & \sqrt{2} & 1\\
\hline
\end{array}$
\hspace{1cm}
$\begin{array}{l|l|l|l|l}
\multicolumn{5}{l}{\ZZ_{2}~toric~code} \\
\hline
 c= 0 & 1&  e & m & em \\
 \hline
 h_i & 0 & 0& 0& 1/2 \\
 d_i & 1 & 1 & 1 &1 \\
\hline
\end{array}$
\end{center}
\caption{Ising and toric code, spins and quantum dimensions.}
\label{tab:ising}
\end{table}%

We wish to consider a situation with an island in the Abelian phase surrounded by a medium in the Ising phase. To achieve this, we start with a large disc in the Ising phase (phase I) and imagine placing a small disk on top of it, which is in a suitable phase III so that a bose condensate can form leaving the bulk of the small disc in the $\mathbb{Z}_2 \times \mathbb{Z}_2$ phase (phase II). Not surprisingly one should use an opposite chirality Ising model for the small phase III disc and then condense the $(\psi,\psi)$ field. This example has been worked out in detail in section X of \cite{Bais2008-08}. Condensation leads to the identifications $(1,1)\sim(\psi,\psi)$, $(\psi,1)\sim(1,\psi)$, $(\sigma,1)\sim(\sigma,\psi)$ and $(1,\sigma)\sim(\psi,\sigma)$, while the remaining field has to split: $(\sigma,\sigma)=(\sigma,\sigma)_1 + (\sigma,\sigma)_2$. Hence, the T-algebra has $6$ sectors and one finds that it has $Ising\times \ZZ_2$ fusion rules. The fields $(\sigma,1)$ and $(1,\sigma)$ will be confined because they cannot be assigned a consistent conformal weight (the corresponding identified $Ising\times\overline{Ising}$ fields have conformal weights that differ by $1/2$). The unconfined fields $(1,1)$, $(\sigma,\sigma)_1$, $(\sigma,\sigma)_2$ and $(\psi,1)$ correspond precisely to the fields $1,e,m$ and $em$ of the $\mathbb{Z}_2 \times \mathbb{Z}_2$ given in table \ref{tab:ising}.

Let us now look at the wall in between the phases. Of the three unconfined fields that correspond to particle excitations in the interior bulk, the fermionic $(\psi,1)$ excitation can freely move out through the wall into the exterior region and  the other two bulk excitations cannot. This corresponds well to the results of \cite{Kells08} where it was shown that free fermionic excitations occur throughout the phase diagram. The confined excitations are expelled from the interior. One, the $(\sigma,1)$ excitation can move into the exterior region, while the other, the $(1,\sigma)$ excitation is strictly confined to the wall. Now consider a $\sigma$ excitation hitting the boundary. From the $T$ theory's fusion rules, we see that $(\sigma,1)=(1,\sigma)\times(\sigma,\sigma)_{1}=(1,\sigma)\times(\sigma,\sigma)_{2}$. Hence, the $\sigma$-particle can split into a boundary excitation and either an $e$ or an $m$ type toric code excitation. This corresponds well with the results of \cite{Wootton08}, where $\sigma$-like excitations were exhibited in the toric code using superpositions of $e$ and $m$ type excitations. Pushing another $\sigma$ particle through the phase boundary will allow the confined $(1,\sigma)$ excitations to annihilate, yielding either $(1,1)$ or $(\psi,1)$. If the two $\sigma$ particles were pair created (had trivial fusion channel), the two toric code particles that form must have fusion channel $1$ resp.~$em\equiv(\psi,1)$, conserving overall $T$-charge.  

\section{The Pfaffian/NASS interface}

Now we turn to the interface between the Moore Read (MR) Pfaffian fractional quantum Hall state at filling $\nu=1/2$ or $5/2$~\cite{mooreread} and the non-abelian spin-singlet (NASS) state of Ardonne and Schoutens at $\nu=4/7$ or $18/7$~\cite{eddyenkjs}. This was recently considered in \cite{Grosfeld2008}. We will again realize it as single layer--two layer boundary. We concentrate on the non-Abelian parts of the the MR and NASS theories here and leave out the $U(1)$ factors (these can be put back in at any point). Consider a  disc with $C_1= Ising$, corresponding to MR, with on top of that a smaller disc with $C_3= M(4,5)$. The latter CFT is the minimal model  with $c=7/10$ corresponding to a tri-critical Ising model. We give the field content of the $Ising$ and $M(4,5)$ theories in tables \ref{tab:ising} and \ref{tab:M4-5}. For the $M(4,5)$ fusion rules we refer to the literature \cite{dsm}.

\begin{table}[htb]
\begin{center}
$\begin{array}{l|l|l|l|l|l|l}
\multicolumn{7}{l}{M(4,5)}\\
\hline
c= 7/10& 1 & \epsilon & \epsilon' & \epsilon'' & \bar{\sigma}& \bar{\sigma}' \\
 \hline
 h_i & 0 & \frac{1}{10} & \frac{3}{5} & \frac{3}{2} & \frac{3}{80} & \frac{7}{16}\\
 d_i & 1 & \frac{1+\sqrt{5}}{2}& \frac{1+\sqrt{5}}{2}& 1& \frac{1+\sqrt{5}}{\sqrt{2}}&\sqrt{2} \\
\hline
\end{array}$
\end{center}
\caption{The tri-critical Ising model $M(4,5)$ (Phase III).}
\label{tab:M4-5}
\end{table}%

The $(\psi,\epsilon'')$ current is the only bosonic channel in the $(Ising \otimes M(4,5))$ model and we assume that it condenses. As this is a simple current it is straightforward to see what happens to the various fields in the model. This is shown in table~\ref{tab:walls}. We start with $3 \times 6 = 18$ fields. In the second row we indicate how $16$ of these fields become pairwise identified (because they are equivalent modulo fusion with $(\psi,\epsilon'')$), and how the other two fields split. 
We are thus left with the $8+4 = 12$ fields of the T-theory, which form an associative fusion algebra. A detailed analysis of the arguments for this case shows that the fusion rules of the algebra are given by $T= M(4,5)\otimes \ZZ_2$ where the nontrivial $\ZZ_2$ element corresponds to the either the $\psi_1$ or the $\psi_2$ field. 
The $T$-fields that are not confined form the residual bulk theory (phase II) and correspond to the fields of the NASS state of \cite{eddyenkjs}, as we intended. Their quantum dimensions and topological spins are given in table \ref{tab:NASS}).
\begin{table}[htb]
\begin{center}
$\begin{array}{l|l|l|l|l|l|l|l|l}
\multicolumn{9}{l}{NASS -state}\\
\hline
c= 6/5& 1 &  \sigma_ \uparrow& \sigma_\downarrow & \sigma_3 & \rho & \psi_1 & \psi_2 & \psi_{12} \\
 \hline
 h_i & 0 & \frac{1}{10} & \frac{1}{10} &\frac{1}{10} &\frac{3}{5} & \frac{1}{2} & \frac{1}{2} & \frac{1}{2} \\
 d_i & 1 & \frac{1+\sqrt{5}}{2}& \frac{1+\sqrt{5}}{2}&\frac{1+\sqrt{5}}{2}&\frac{1+\sqrt{5}}{2}& 1& 1  &1    \\
\hline
\end{array}$
\end{center}
\caption{The NASS state (Phase II).}
\label{tab:NASS}
\end{table}
The twelve fields of the full T-theory describing the interface are listed at the top of table \ref{tab:walls}. We also give their quantum dimensions, which are consistent with the fusion rules and with the decomposition of the original sectors. It is not possible to assign unambiguous conformal weights to the confined fields, and therefore there is no consistent braid group representation for the full T-theory. We recall (cf.~\cite{baismathy06b}) that this is no problem, since the full T-theory has only a strictly one-dimensional interpretation. 

It is now interesting to look more closely at the properties at the walls between the phases. In table \ref{tab:walls} we have explicitly indicated the fields in the different possible interior and exterior regions, as well as those being strictly confined to particular walls, where the latter explicitly depend on what wall we are talking about. More precisely, 
we give the fields that extend from the wall into the exterior MR (phase I) region in the fourth row and the fields that extend from the wall into the interior NASS (phase II) region in the fifth row, and we have to conclude that the three remaining fields can only propagate in the wall. Note that we also have to identify the $\psi_{12}$ field in the NASS region, with the $\psi$ field in the MR region, which means that this field can propagate right through the wall. 

It is  clear that the fusion rules of the T-theory  fix the kinematically allowed channels in which particles which hit the wall, coming from either the interior or the exterior region, can split into particles in the other region plus a wall-excitation. For instance, from the T fusion rule $\sigma \times \bar{\sigma} = \sigma_\uparrow+ \sigma_\downarrow$, we find that if we start with a $\sigma_\uparrow$ coming from the interior region, it can split into a $\sigma$ going into the MR region and a $\bar{\sigma}$ staying in the wall. But since $\bar{\sigma} \times \sigma^* = \sigma_\uparrow + \sigma_\downarrow+ \psi_1 + \psi_2$, it may also happen that the $\sigma_\uparrow$ excitation splits into the two wall-excitations  $\bar{\sigma}$ and ${\sigma^*}$. This scenario may also be turned around, two strict boundary excitations may fuse into a state that is not confined. Obviously there is a myriad of possibilities and we refrain from listing them here. 

A final comment concerns the relaxation of qubits near a wall \cite{Ilan2008}. If we encode a topological qubit in the NASS phase, for example in a pair of sigma particles (which indeed span a two-dimensional Hilbert space) the qubit may relax to the lowest energy state by transferring  a neutral excitation to the boundary.  For a pair of $\sigma_3$ fields we have the fusion rule $\sigma_3 \times \sigma_3=1+ \rho$, while $\sigma_\downarrow \times \sigma_\uparrow$ = $\psi_{12}+ \sigma_3$. Each of these pairs can relax under emission of a $\rho$ excitation. A $\rho$ excitation may convert into one of the pairs  $(\sigma^*,\sigma^*)$, $(\bar{\sigma},\bar{\sigma})$, or  $(\bar{\sigma},\bar{\sigma}')$, which are all strictly confined to the interface. Alternatively we may have $\rho \rightarrow (\sigma^*,\sigma)$ where $\sigma^*$ is confined to the wall but the $\sigma$ can enter in the MR region.

\acknowledgments{We thank Prof.~K.~Schoutens for useful discussions.}

\begin{widetext}

\begin{table}[tbh]
\begin{center}
$\begin{array}{l|l|l|l|l|l|l|l|l|l|l|l|l}
\hline
T-theory & 1 &  \sigma_ \uparrow& \sigma_\downarrow & \sigma_3 & \rho & \psi_1 & \psi_2 & \psi_{12} &\sigma & \bar{\sigma} & \bar{\sigma}' & \sigma^*\\
 \hline
Corresponding~sectors&(1,1) & \multicolumn{2}{c|}{(\sigma,\bar{\sigma})}&  (1,\epsilon)  & (1,\epsilon')& \multicolumn{2}{c|}{(\sigma,\bar{\sigma}')}&(1,\epsilon'') & (\sigma,1)&(1,\bar{\sigma})& (1,\bar{\sigma}')& (\sigma, \epsilon)  \\
in~M(4,5)\otimes Ising& (\psi, \epsilon'') & \multicolumn{2}{c|}{}&(\psi,\epsilon') &(\psi,\epsilon)  & \multicolumn{2}{c|}{}&(\psi,1) & (\sigma,\epsilon'') &(\psi,\bar{\sigma}) & (\psi,\bar{\sigma}') & (\sigma,\epsilon')\\ \hline
 d_i & 1 & \frac{1+\sqrt{5}}{2} & \frac{1+\sqrt{5}}{2}&\frac{1+\sqrt{5}}{2}&\frac{1+\sqrt{5}}{2}& 1 & 1  &1& \sqrt{2}&\frac{1+\sqrt{5}}{\sqrt{2}}&\sqrt{2}&\frac{1+\sqrt{5}}{\sqrt{2}}   \\
 \hline
Phase~I: MR & 1 & & & & & & &\psi & \sigma & & \\
Phase~II: NASS &1 &  \sigma_ \uparrow& \sigma_\downarrow & \sigma_3 & \rho & \psi_1 & \psi_2 & \psi_{12} & & & \\
Confined~on~I/II~wall& & & & & & & & & & \bar{\sigma} & \bar{\sigma}' & \sigma^* \\
Phase~III: M(4,5) & 1 & & & \epsilon & \epsilon' & & & \epsilon'' & &  \bar{\sigma} & \bar{\sigma}' & \\
Confined~on~II/III~wall& & & & & & & & &\sigma &  & & \sigma^* \\
\hline
\end{array}$\\
\end{center}
\caption{Field content of the T-theory resulting from a $(\psi, \epsilon'')$ condensate in the $M(4,5)\otimes Ising$ model and governing the kinematics of the NASS and MR states and the domain wall between them. The following rows give the correspondence between T-sectors and sectors of the different phases and walls. One reads off that the fields $\bar{\sigma}, \bar{\sigma}' $ and $\sigma^*$ are strictly confined to the I-II boundary. The same T-theory would live on a domain wall between NASS and M(4,5) phases, where the fields $\sigma$ and $\sigma^*$ would be strictly confined to the II/III boundary. Clearly on an edge between the NASS phase and the vacuum, one would find the confined fields $\sigma, \bar{\sigma}, \bar{\sigma}'$ and $\sigma^*$.}  
\label{tab:walls}
\end{table}

\end{widetext}


\end{document}